\documentclass{article}

\usepackage[english]{babel}

\usepackage[letterpaper,top=2cm,bottom=2cm,left=3cm,right=3cm,marginparwidth=1.75cm]{geometry}

\usepackage{natbib}
\usepackage{amsmath}
\usepackage{amssymb}
\usepackage{amsfonts}
\usepackage{graphicx}
\usepackage{bm}
\usepackage[colorlinks=true, allcolors=blue]{hyperref}
\usepackage{doi}

\newcommand{\xf}{\mathbf{x}^\text{f}}
\newcommand{\xa}{\mathbf{x}^\text{a}}
\newcommand{\xb}{\mathbf{x}^\text{b}}
\newcommand{\xt}{\mathbf{x}^\text{t}}
\newcommand{\y}{\mathbf{y}}

\newcommand{\Pf}{\mathbf{P}^\text{f}}
\newcommand{\K}{\mathbf{K}}
\newcommand{\R}{\mathbf{R}}
\newcommand{\h}{\mathbf{H}}
\newcommand{\Mml}{M^\text{ML}}

\DeclareMathOperator*{\argmax}{arg\,max}
\DeclareMathOperator*{\argmin}{arg\,min}

\title{The interface of data assimilation and machine learning}
\date{5 October 2026}
\author{Eviatar Bach}

\begin{document}
\maketitle

\begin{abstract}
\noindent A slightly older version of this article will appear in the book \emph{Weather and Climate: Applications of Machine Learning and Artificial Intelligence} by Driscoll et al. (2027). If you would like to cite this, please use
\vspace{10pt}

\noindent Bach, E. (2027). The Interface of Data Assimilation and Machine Learning. In S. Driscoll, K. M. R. Hunt, L. Mansfield, R. Swaminathan, H. Wei, E. Bach, \& A. Peard, \emph{Weather and Climate: Applications of Machine Learning and Artificial Intelligence}. Elsevier.
\end{abstract}

\section{Introduction}

Data assimilation (DA) is the process of combining forecasts from a model with observations in order to optimally estimate the state of a system. This is critical for chaotic systems, such as the atmosphere, since if observations are not continually assimilated the model will quickly lose skill. DA is routinely performed (usually every 6 hours) at operational forecasting centres around the world.

In this chapter we will discuss the interface of machine learning (ML) and DA. This is still an emerging and quickly developing field, and this chapter tries to give an overview of some of the main topics and methods.

We only overview methods in a cursory way, with pointers to the wider literature. A more mathematical and comprehensive overview of the topics discussed here is provided in \cite{bach_inverse_2024}, and some of the exposition here is based on the latter. See also the review papers of \cite{cheng_machine_2023,bocquet_surrogate_2023}.

\section{Data assimilation fundamentals}\label{sec:da}

A classic textbook on DA from a meteorological perspective is \cite{kalnay_atmospheric_2002}, and is still one of the best books on the subject. Other recommended references that take a more mathematical perspective are \cite{law_data_2015,asch_data_2016,sanz-alonso_inverse_2023}. A good review paper on DA in the geosciences is \cite{carrassi_data_2018}. We use here the notation of \cite{ide_unified_1997}, which is common in the meteorological and oceanographic literature.

An important topic in DA is that of uncertainty quantification, which is naturally expressed in a Bayesian framework. Within this framework DA is known as \emph{filtering} and \emph{smoothing}. In this chapter we will generally focus on the problem of \emph{state estimation}, which instead tries to obtain a point estimate. However, many of the methods discussed below can be extended to the Bayesian problem of estimating the posterior, and we will point to further resources when this is the case. A unified Bayesian perspective on many of these topics is taken in \cite{bach_inverse_2024}.

We begin by considering a system that evolves according to a hidden true state $\xt$ in time according to
\begin{equation}\label{eq:dynamics}
    \xt(t_{i+1}) = M(\xt(t_i)),
\end{equation}
where we assume access to the forecast model $M$.

It is also common to consider a noisy version of the preceding equation, 
\begin{equation}
    \xt(t_{i+1}) = M(\xt(t_i)) + \bm{\eta}(t_i),
\end{equation}
where
\begin{equation}
    \bm{\eta}(t_i) \sim \mathcal{N}(\mathbf{0}, \mathbf{Q}).
\end{equation}
The noise $\bm{\eta}$ is often used to capture the impact of \emph{model error}: a source of error growth due to the deficiencies in our model of the system. Note that model error will typically be biased (not centred at zero) and state-dependent; we will address that case later in the chapter.

Then, we have access to observations obtained by
\begin{equation}\label{eq:obs}
    \mathbf{y}(t_i) = H(\xt(t_i)) + \bm{\varepsilon}(t_i),
\end{equation}
where
\begin{equation}
    \bm{\varepsilon}(t_i)\sim\mathcal{N}(\mathbf{0}, \R).
\end{equation}
$H$ is called the \emph{observation operator}, mapping from the state space of the model to the observation space.

Note that both $M$ and $H$ can vary in time in general, but for simplicity we assume they are time-independent.

The problem of \emph{data assimilation} (DA) is to estimate the true states $\{\xt(t_i)\}_{i=0}^n$ given the model $M$ and observations $\{\y(t_i)\}_{i=0}^n$. DA algorithms alternate between \emph{forecast} and \emph{analysis} steps:
\begin{enumerate}
    \item \emph{Forecast}: Given the analysis state at time $t_i$, use the model $M$ to forecast to the next time $t_{i+1}$.
    \item \emph{Analysis}: Given the forecast state at time $t_{i+1}$, assimilate the observations $\y(t_{i+1})$ to produce the analysis at time $t_{i+1}$.
\end{enumerate}
\begin{figure}
    \centering
    \includegraphics[width=0.65\linewidth]{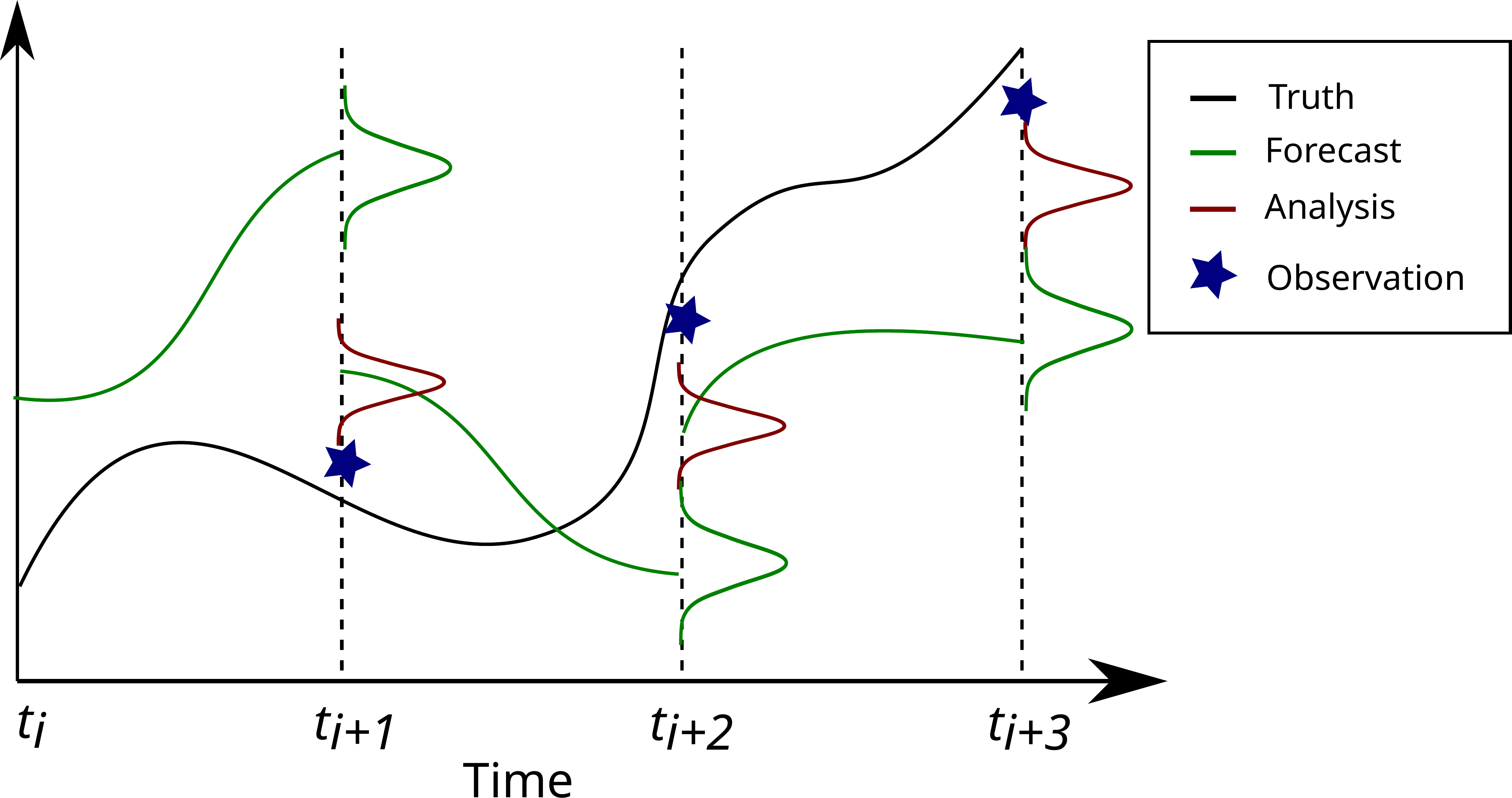}
    \caption{A schematic of the data assimilation cycle. The forecast and analysis are shown as probability distributions, in keeping with the probabilistic formulation of DA.}
    \label{fig:da_cycles}
\end{figure}
The forecast--assimilation cycle is shown in Figure~\ref{fig:da_cycles}. Different DA algorithms differ in how they implement these steps. In the subsections that follow, we will cover three popular DA algorithms: 3DVar, 4DVar, and the ensemble Kalman filter. 3DVar and 4DVar are both known as \emph{variational methods}.

\emph{Reanalysis}, or retrospective analysis, refers to the process of producing a ``best estimate'' of the atmosphere's state in the past, over an extended period of time, by assimilating all possible observations using a consistent numerical weather prediction (NWP) model version. Reanalysis datasets are often used in studying climate and, as we shall see, play an important role in the training of ML models.

\subsection{3DVar}



Suppose that in addition to the observation $\y(t_i)$, we have at $t_i$ a ``first guess'' or prior for the state estimate. This is known as the \emph{background}, and denoted by $\xb(t_i)$. We take this background to be distributed as
\begin{equation*}
    \xb(t_i)\sim \mathcal{N}(\xt(t_i), \mathbf{B}(t_i));
\end{equation*}
that is, the background has error covariance matrix $\mathbf{B}$.

An analysis $\xa(t_i)$, given the observations and background, is obtained by minimising the cost function 
\begin{subequations}\label{eq:3dvar}
\begin{align}
    J_i(\mathbf{x}) &= \frac{1}{2}(\mathbf{x} - \xb(t_i))^\top \mathbf{B}(t_i)^{-1}(\mathbf{x} - \xb(t_i)) + \frac{1}{2}(\y(t_i) - H(\mathbf{x}))^\top\R^{-1}(\y(t_i) - H(\mathbf{x})),\\
    \xa(t_i) &= \argmin_\mathbf{x} J_i(\mathbf{x}).
\end{align}
\end{subequations}
This is called the \emph{3DVar} (3-dimensional variational) method. For a linear observation operator $H = \h$, the solution can be explicitly written as
\begin{equation}\label{eq:oi}
    \xa(t_i) = \xb(t_i) + \mathbf{B}(t_i)\h^\top(\h \mathbf{B}(t_i)\h^\top + \R)^{-1}(\y(t_i) - \h\xb(t_i));
\end{equation}
this form is referred to as \emph{optimal interpolation}.

Typically, 3DVar is performed sequentially, such that the background at time $t_{i+1}$ is obtained by applying the forecast model to the last analysis:
\begin{equation*}
    \xb(t_{i+1}) = M(\xa(t_i)).
\end{equation*}
One thus only needs to specify a background at time $t_0$, $\xb(t_0)$. In this setting, the background $\xb$ can be denoted as the \emph{forecast} $\xf$.

While simple, 3DVar has the disadvantage of needing to specify the background covariance matrix $\mathbf{B}$. Methods to do this are discussed in \cite{kalnay_atmospheric_2002}. Moreover, $\mathbf{B}$ is often taken to be fixed in time, meaning that the background uncertainty cannot vary dynamically. Dynamic forecast covariance is one of the key advantages of the ensemble Kalman filter, discussed below.

\subsection{4DVar}

\subsubsection{Strong constraint}
The \emph{4DVar} (4-dimensional variational) method, specifically the \emph{strong constraint} variant, extends 3DVar over multiple observation times $t_0, \ldots, t_n$ using the following cost function:
\begin{subequations}\label{eq:4dvar}
\begin{align}
    J(\mathbf{x}(t_0)) =& \frac{1}{2}(\mathbf{x}(t_0) - \xb(t_0))^\top \mathbf{B}(t_0)^{-1}(\mathbf{x}(t_0) - \xb(t_0))\\
    &\qquad+ \frac{1}{2}\sum_{i=0}^n(\y(t_i) - H(\mathbf{x}(t_i)))^\top\R^{-1}(\y(t_i) - H(\mathbf{x}(t_i))),\nonumber\\
    \xa(t_0) =& \argmin_\mathbf{x} J(\mathbf{x}).
\end{align}
\end{subequations}
The optimisation is over the initial condition $\mathbf{x}(t_0)$ since
\begin{equation*}
    \mathbf{x}(t_i) = (\underbrace{M\circ\cdots\circ M}_\text{$i$ times})(\mathbf{x}(t_0)),
\end{equation*}
where $\circ$ indicates function composition. The trajectory $\mathbf{x}(t_0),\ldots,\mathbf{x}(t_n)$ is therefore constrained to be a trajectory of the model; thus the name ``strong constraint''.

4DVar, as with 3DVar, can also be applied in a sequential way. This is often done by splitting the times $t_0,\ldots,t_n$ into shorter blocks, applying 4DVar to each of these, taking the background of one block to be the analysis at the end of the previous block.

The typical method for optimising the 4DVar cost function involves the \emph{linear tangent model}, a linearised version of the dynamics. Obtaining and maintaining the linear tangent model for an NWP model often requires a great deal of human effort; avoiding this difficulty is an appeal for the use of ML discussed below.

\subsubsection{Weak constraint}

We can account for model error by adding an extra term to the cost function. This is called \emph{weak constraint 4DVar}:
\begin{subequations}
\begin{align}
    J(\mathbf{x}(t_0),\ldots,\mathbf{x}(t_n)) =& \frac{1}{2}(\mathbf{x}(t_0) - \xb(t_0))^\top \mathbf{B}(t_0)^{-1}(\mathbf{x}(t_0) - \xb(t_0))\\
    &\qquad+ \frac{1}{2}\sum_{i=0}^n(\y(t_i) - H(\mathbf{x}(t_i)))^\top\R^{-1}(\y(t_i) - H(\mathbf{x}(t_i)))\nonumber\\
    &\qquad+ \frac{1}{2}\sum_{i=1}^n (\mathbf{x}(t_i) - \mathcal{M}(\mathbf{x}(t_{i-1})))^\top\mathbf{Q}^{-1}(\mathbf{x}(t_i) - \mathcal{M}(\mathbf{x}(t_{i-1}))),\nonumber\\
    \xa(t_0), \ldots, \xa(t_n) =& \argmin_{\mathbf{x}(t_0),\ldots,\mathbf{x}(t_n)} J(\mathbf{x}(t_0),\ldots,\mathbf{x}(t_n)).
\end{align}
\end{subequations}
Note that here the solution to the optimisation is no longer constrained to be a trajectory of the model, thus the name ``weak constraint''. Hence, the optimisation is no longer over just the initial condition, rather the whole trajectory.


\subsection{Ensemble Kalman filter}

The \emph{ensemble Kalman filter (EnKF)} takes a different approach than the variational methods discussed above. The EnKF evolves an ensemble of states to represent the estimated uncertainty at a given time. This dynamic uncertainty is a key advantage over variational methods. In the forecast step, the model is applied to each ensemble member, obtaining a forecast ensemble $\{\xf_j\}_{j=1}^m$, and in the analysis step the observations are assimilated, resulting in an analysis ensemble $\{\xa_j\}_{j=1}^m$.

The most basic form of the EnKF is the stochastic EnKF which, in the case that the observation operator is linear ($H = \h$), is given by
\begin{subequations}
\begin{align}
    \xf_j(t_{i+1}) &= M(\xa_j(t_i)) + \bm{\eta}_j(t_i),\qquad\bm{\eta}_j(t_i) \sim \mathcal{N}(\mathbf{0}, \mathbf{Q}),\\
    \xa_j(t_{i+1}) &=  \xf_j(t_{i+1}) + \K(t_i)(\y(t_i) + \bm{\varepsilon}_j(t_i) - \h \xf_j(t_{i+1})),\quad\bm{\varepsilon}_j(t_i) \sim \mathcal{N}(\mathbf{0}, \R),\label{eq:enkf_analysis}\\
    \mathbf{K}(t_i) &= \Pf(t_i) \h^\top(\h \Pf(t_i)\h^\top + \R)^{-1},\\
    \overline{\mathbf{x}}^\text{f}(t_i) &= \frac{1}{m} \sum_{j=1}^m \xf_j(t_i),\\
    \overline{\mathbf{x}}^\text{a}(t_i) &= \frac{1}{m} \sum_{j=1}^m \xa_j(t_i),\\
    \Pf(t_i) &= \frac{1}{m}\sum_{j=1}^m (\xf_j(t_i) - \overline{\mathbf{x}}^\text{f}(t_i))(\xf_j(t_i) - \overline{\mathbf{x}}^\text{f}(t_i))^\top,
\end{align}
\end{subequations}
where an initial forecast ensemble $\{\xf_j(t_0)\}_{j=1}^m$ must also be specified. At each time $t_i$ the analysis ensemble mean $\overline{\mathbf{x}}^\text{a}(t_i)$ is taken to be the estimate of the system state.

Note the similarity between \eqref{eq:enkf_analysis} and the optimal interpolation solution of 3DVar \eqref{eq:oi}. The EnKF can be thought of as applying 3DVar to each ensemble member, using an empirical covariance matrix $\Pf$ rather than a background covariance matrix $\mathbf{B}$.

The EnKF can also be applied for nonlinear observation operators by replacing $\Pf\h^\top$ by the sample cross-covariance between the ensemble and the ensemble in observation space, and $\h\Pf\h^\top$ by the covariance of the ensemble in observation space.

For high-dimensional systems, the EnKF generally needs two modifications to cope with sampling error: inflation and localisation. \emph{Inflation} increases the forecast covariance matrix by a factor $\alpha > 1$:
\begin{equation*}
    \Pf(t_i) \to \alpha \Pf(t_i).
\end{equation*}
\emph{Localisation} modifies $\Pf$ such that correlations between locations far from each other in physical space are damped, in order to lessen the impact of spurious long-range correlations \citep{carrassi_data_2018}. Furthermore, EnKFs are typically implemented as ensemble square-root filters, in which a covariance matrix is not explicitly formed \citep{tippett_ensemble_2003}. A popular implementation of the EnKF, which is both a square-root filter and implements localisation, is the local ensemble transform Kalman filter \cite[LETKF,][]{hunt_efficient_2007}.

Variational and ensemble algorithms have their strengths and weaknesses \citep{kalnay_4-d-var_2007}, and are both widely used. \emph{Hybrid ensemble--variational methods} combine the strengths of both approaches, and usually have the best performance \citep{carrassi_data_2018}.

\section{Machine learning for data assimilation}

In what follows, we will consider the training of ML algorithms abstractly as an optimisation problem for parameters $\boldsymbol\theta$. In the presentation of the algorithms, we will not discuss specific ML architectures in order to maintain generality. $\boldsymbol\theta$ could represent, for instance, the weights and biases of a neural network.

We also avoid discussing specific optimisation methods, instead presenting cost functions $J(\boldsymbol\theta)$ to be minimised with respect to $\boldsymbol\theta$. In many of these cases, optimisation will require \emph{automatic differentiation} to compute gradients of the cost function. Automatic differentiation is an important component of many ML software libraries. In some cases discussed below, the methods will require the gradients of the forecast model, the observation operator, and/or the DA algorithm, which may pose practical difficulties for implementation.

\subsection{Using data assimilation to learn a model}

Probably the most widespread application of the combination of ML and DA is in learning an ML forecast model, or learning a model error correction to an existing physics-based model.

\subsubsection{Learning a forecast model using reanalysis}\label{ssec:reanalysis}
Reanalysis is widely used to train ML forecast models. This is, indeed, how most of the ML weather models discussed in Chapter 4 were trained. Why train on reanalysis? Consider the alternatives:
\begin{enumerate}
    \item \emph{Why not train directly on observations?} Observations are typically sparse and noisy, and irregular in space and time. Reanalysis functions as a physics-based interpolator, being generally more accurate than the observations themselves, and available at every point on the model grid at regular time intervals.
    \item \emph{Why not train directly on model output?} If we train on physics-based model output, we cannot do better than the model itself, and would inherit any inaccuracies it might have.
\end{enumerate}
We add two caveats to these points. Although reanalysis will generally have less model bias than raw model output, it will still have some remaining bias. This has led to attempts to train directly on observations (see section~\ref{sec:end-to-end}). On the other hand, training on model output can be useful if one is looking for a computationally cheap surrogate rather than trying to capture as faithfully as possible the system's true dynamics.

Consider a reanalysis dataset $\{\xa(t_i)\}_{i=0}^n$, defined for times $t_0,\ldots,t_n$. Then, we would like to train a one-step ML prediction model, $\Mml$, which predicts the state at time $t_{i+1}$ when given as input the state at time $t_i$. This model has parameters $\boldsymbol\theta$. We can use a cost function of the form
\begin{equation}\label{eq:mse_cost}
    J(\boldsymbol\theta) = \frac{1}{n}\sum_{i=0}^{n-1} \|\xa(t_{i+1}) - \Mml(\xa(t_i); \boldsymbol\theta)\|^2.
\end{equation}

It is often useful to include multi-step terms in the cost function; i.e., to penalise $\|\xa(t_{i+2}) - \Mml(\Mml(\xa(t_i)))\|^2$, $\|\xa(t_{i+3}) - \Mml(\Mml(\Mml(\xa(t_i))))\|^2$, etc. This technique, known as \emph{rollout}, helps the ML forecast model to have better long-term behaviour and stability, but comes with additional training cost.

Recent approaches have made use of probabilistic scoring rule--based cost functions rather than the mean-square one in \eqref{eq:mse_cost}. The training data is still reanalysis, but an ensemble of ML predictions is used as the forecast, and the scoring rule rewards accurate probabilistic performance. A common choice of scoring rule is the continuous ranked probability score (CRPS), which has the appealing theoretical property of being \emph{strictly proper} (see also Chapter 6 for more on the CRPS). An example of a (hybrid) ML model trained in this way is NeuralGCM \citep{kochkov_neural_2024}. See \citet[][Chapter 11]{bach_inverse_2024} for more information on scoring rules and their use in learning.

\subsubsection{Learning a model error correction using analysis increments}

\begin{figure}
    \centering
    \includegraphics[width=0.65\linewidth]{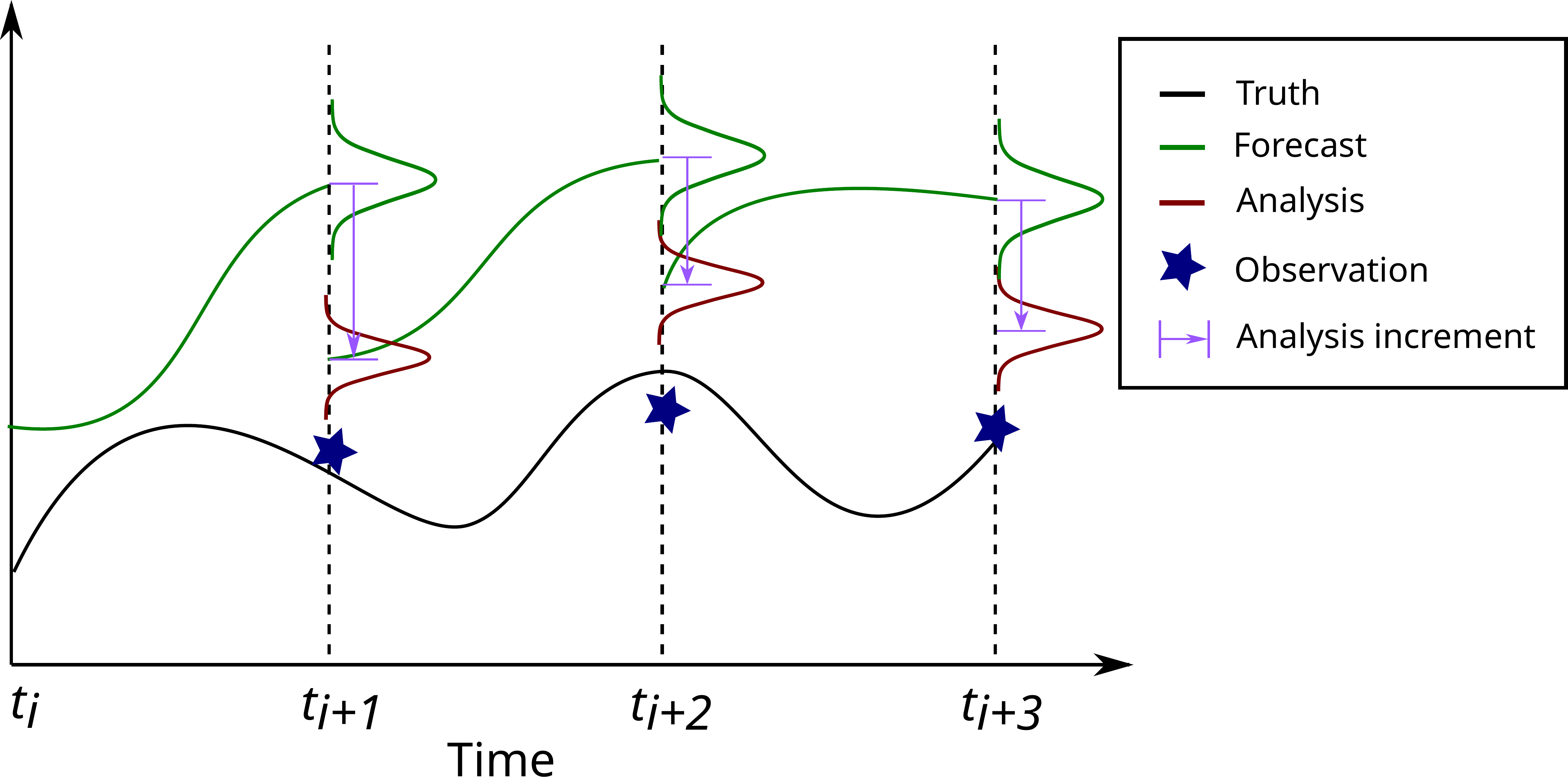}
    \caption{A schematic showing analysis increments in the presence of a systematic model bias. In this case, the forecast model is biased upwards, and the analysis increments point in the opposite direction. Note that the analysis still retains some of the bias.}
    \label{fig:analysis_increments}
\end{figure}
Another way to use reanalysis is to learn model error corrections to improve an imperfect physics-based model. As noted above, reanalysis generally has less model bias than raw model output, since the model forecast is ``nudged'' towards the observations, which are usually assumed to be unbiased but noisy. These nudges towards the observations, that is $\text{AI}(t_i) = \xa(t_i) - \xf(t_i)$, are called \emph{analysis increments}. A simple schematic is shown in Figure~\ref{fig:analysis_increments}.

The idea of this method \citep{danforth_estimating_2007,farchi_using_2021} is to learn systematic patterns in past analysis increments to improve the model, even when observations are not available. The whole analysis increment will not be predictable without the observations, but some patterns might be.

For instance, if the analysis consistently pushes the forecast in a given direction, to counteract a model bias, the average of the past analysis increments will reflect this bias, and we can improve a forecast $\xf(t_i)$ by adding this averaged analysis increment:
\begin{equation*}
    \xf(t_i) \to \xf(t_i) + \overline{\text{AI}},
\end{equation*}
where $\overline{\text{AI}}$ is the average of past analysis increments. More generally, an ML model $\text{AI}^\text{ML}$ can be trained to predict the analysis increment as a function of the forecast state; this allows one to potentially capture state-dependent biases. A cost function
\begin{equation*}
    J(\boldsymbol\theta) = \frac{1}{n+1}\sum_{i=0}^n\|\text{AI}(t_i) - \text{AI}^\text{ML}(\xf(t_i); \boldsymbol\theta)\|^2.
\end{equation*}
After optimising for $\boldsymbol\theta$, the optimal ${\boldsymbol\theta}^*$ can be used for future times $t_j$,
\begin{equation*}
    \xf(t_j) \to \xf(t_j) + \text{AI}^\text{ML}(\xf(t_j); \boldsymbol\theta^*).
\end{equation*}

Such methods have been successfully applied in large-scale NWP settings \citep{chen_correcting_2022,farchi_development_2025}.

\subsubsection{Online methods}

The previous two methods relied on there being a database of past analysis states produced using a physics-based model. These methods are called \emph{offline}, in the sense that DA does not need to be applied during the training process once this database is available. This is advantageous in terms of computational efficiency, since performing DA repeatedly is typically very expensive. The disadvantage of these methods is that analysis/reanalysis does not totally remove the impact of model error, as previously discussed.

An alternative class of \emph{online} methods incorporate data assimilation into the training process \citep[see][Chapter 8 for an overview]{bach_inverse_2024}. These methods can be used both for learning a forecast model, learning a model error correction, or learning an unknown model error covariance $\mathbf{Q}$ (for more on the latter see \cite{tandeo_review_2020}). They are typically based on \emph{maximum likelihood}: given an observation dataset $\{\y(t_i)\}_{i=0}^n$, we would like to choose the parameters that maximise the likelihood of the observations:
\begin{equation*}
    {\boldsymbol\theta}^* = \argmax_{\boldsymbol\theta} \mathbb{P}(\{\y(t_i)\}_{i=0}^n | \boldsymbol\theta).
\end{equation*}

Using rules of conditional probability, we can rewrite the likelihood
\begin{equation*}
    \mathbb{P}(\{\y(t_i)\}_{i=0}^n | \boldsymbol\theta) = \int \mathbb{P} (\{\y(t_i)\}_{i=0}^n | \{\xt(t_i)\}_{i=0}^n, \boldsymbol\theta) \mathbb{P} (\{\xt(t_i)\}_{i=0}^n | \boldsymbol\theta) \, d\xt(t_0)\ldots d\xt(t_n).
\end{equation*}
This demonstrates the difficulty of the optimisation problem posed above: in order to estimate the likelihood, we must first estimate the underlying states $\{\xt(t_i)\}_{i=0}^n$, which is the task of DA.

We will not go into further details here, but online methods typically work by iteratively applying training and assimilation steps. Methods to do this include ones based on expectation--maximisation \citep{bocquet_bayesian_2020} and the autodifferentiable EnKF \citep{chen_autodifferentiable_2022}. Such online methods are much more expensive than the offline ones, but they are not limited by the errors in analysis. To our knowledge, such methods have not yet been applied in operational NWP settings.

Another approach is to perform DA with an augmented state vector that also includes the ML parameters $\boldsymbol\theta$ \citep{gottwald_combining_2021,gottwald_supervised_2021}. More recently, the European Centre for Medium-Range Weather Forecasts (ECMWF) is testing a system wherein an offline training stage is followed by online tuning with DA \citep{farchi_development_2025}.

\subsection{Applying data assimilation to a machine learning model}

Once an ML model is trained, in order to use it for forecasting, we must apply DA to it. Ostensibly, this involves replacing $M$ with $\Mml$ in any DA algorithm, such as those discussed in Section~\ref{sec:da}.

If the ML model is a close enough approximation to the true system dynamics, i.e., $\Mml\approx M$, then it can be shown that the estimated trajectory will stay close to the hypothetical trajectory estimated using the true dynamics. Furthermore, even if the ML model is unstable for long rollouts, a common problem, frequent DA can prevent blow-up \citep{adrian_data_2025}. See Chapter 6 in \citet{bach_inverse_2024} for further theoretical results.

In practice, it is important for DA that the ML model accurately capture the error growth characteristics of the system, as well as the covariance structure between variables. \cite{platt_constraining_2023} consider enforcing the former during the training process.

Application of DA to global ML atmospheric models has thus far yielded mixed results. In experiments with synthetic observations, \cite{adrian_data_2025,li_fuxi-en4dvar_2024,kotsuki_ensemble_2025} achieved reasonable results, although it was found that the ML models did not always accurately capture error covariances. Experiments with real surface pressure observations proved less successful, with poor estimation of cross-variable covariances \citep{slivinski_assimilating_2025}. Other works with more simplified models have also demonstrated the potential of DA with ML forecast models \citep{chattopadhyay_towards_2022,penny_integrating_2022}.

The use of ML forecast models in DA has several potential advantages compared to the use of physics-based models. Reduced computational expense is an attractive prospect, since DA is among the most computationally expensive tasks in operational weather forecasting centres. Additionally, ML forecast models may offer other advantages for variational and ensemble methods, as we now outline.

A potential advantage of using ML models with 4DVar is that, since these models are usually implemented using ML libraries that allow for automatic differentiation, the linear tangent model can be obtained relatively easily, or the 4DVar cost function differentiated directly \citep{tian_evaluating_2026,solvik_4d-var_2025}. However, \cite{tian_evaluating_2026} notes that the linear tangent model of ML models can exhibit unphysical noise that may cause robustness issues.

A potential appeal of using ML models for ensemble DA, on the other hand, is that due to their being typically computationally cheaper to run once trained, one can potentially afford much larger ensembles than otherwise feasible with physics-based models. Such large ensembles, if they accurately captured the time-evolving probability distribution of the system, could potentially improve ensemble DA. Methods that combine physics-based with ML ensembles have also been proposed \citep{chattopadhyay_deep_2023,bach_multi-model_2023,bach_improved_2024,voort_balancing_2025}.

\subsection{Learning the analysis step}

Existing DA methods, while very powerful, typically make simplifying assumptions that make them suboptimal in realistic settings. Most methods are based on Gaussian assumptions that are not always satisfied. A recent and rapidly developing topic is learning the analysis step itself \citep{mccabe_learning_2021,bach_learning_2025,bocquet_accurate_2024,hammoud_data_2024,bach_learning_2026}.

Abstractly, the analysis step of a DA algorithm takes as input a forecast at time $t_i$, $\xf(t_i)$, and an observation at $t_i$, $\y(t_i)$, to produce an analysis $\xa(t_i)$. We can now think of learning a function ${\xa}^{\text{,ML}}(\xf, \y; \boldsymbol\theta)$ with the same inputs. Integrating this ML analysis step into the forecast--analysis cycle and measuring how closely it tracks the true trajectory, we obtain the cost function
\begin{subequations}
\begin{align}
    J(\boldsymbol\theta) &= \frac{1}{n+1}\sum_{i=0}^n\|\xt(t_i) - {\xa}(t_{i}; \boldsymbol\theta)\|^2,\label{eq:obj_analysis}\\
    {\xa}(t_{i}; \boldsymbol\theta) &= {\xa}^{\text{,ML}}(\xf(t_{i}), \y(t_{i}); \boldsymbol\theta),\\
    \xf(t_{i+1}) &= M({\xa}(t_i)).
\end{align}
\end{subequations}
This presupposes that a synthetic truth $\{\xt(t_i)\}_{i=0}^n$, along with simulated observations $\{\y(t_i)\}_{i=0}^n$, have been produced according to \eqref{eq:dynamics} and \eqref{eq:obs}. Note that optimising \eqref{eq:obj_analysis} requires the gradient of the forecast model $M$.

In the case that the available model is highly imperfect, the preceding approach may not be a good one, since the synthetic ``truth'' is not representative of the true system dynamics. To avoid the problem of model error, one can directly train on observations
\begin{subequations}
\begin{align}
    J(\boldsymbol\theta) &= \frac{1}{n+1}\sum_{i=0}^n\|\y(t_i) - H({\xa}(t_i;\boldsymbol\theta))\|^2,\\
    {\xa}(t_{i};\boldsymbol\theta) &= {\xa}^{\text{,ML}}(\xf(t_{i}), \y(t_{i}); \boldsymbol\theta),\\
    \xf(t_{i+1}) &= M({\xa}(t_i)).
\end{align}
\end{subequations}
However, this latter formulation requires care to avoid overfitting to the observations, since they are used as both input and in the cost function; see \cite{mallia-parfitt_assessing_2016}. Optimising this cost function requires the gradient of $H$ in addition to the gradient of $M$.

These learning problems are \emph{amortised}, in the sense that once trained, the learned analysis step can be used in the future for DA without retraining. This relies, however, on the assumption that the model dynamics and observation network do not change in time, and that $n$ is sufficiently large to adequately generalise to unseen data.

Probabilistic methods that attempt to not only match the true state as closely as possible, but also to get correct Bayesian uncertainty quantification, have also been proposed. A framework based on transport maps was introduced in \cite{spantini_coupling_2022}, one based on variational inference in \cite{bach_learning_2025}, and one based on scoring rules in \cite{bachLearningProbabilisticFilters2026}.

Learning the analysis step can be done to attempt to learn a more accurate analysis than that provided by traditional algorithms. A different task is to directly emulate an existing algorithm for reducing computational expense; this can be done using supervised learning with a history of paired forecasts, observations, and analysis states. Alternatively, one can directly use the variational cost functions from 3DVar or 4DVar to train an ML emulator of these algorithms \citep{filoche_learning_2023,keller_ai-based_2024,zinchenko_combined_2024}. For instance, taking the 3DVar cost function \eqref{eq:3dvar} and replacing the minimisation over states by a minimisation over parameters:
\begin{align}
    J(\boldsymbol\theta) =& \frac{1}{2}\left[{\xa}(t_{i}; \boldsymbol\theta) - \xb(t_i)\right]^\top \mathbf{B}(t_i)^{-1}\left[{\xa}(t_{i}; \boldsymbol\theta) - \xb(t_i)\right]\\
    &\qquad+ \frac{1}{2}\left[\y(t_i) - H({\xa}(t_{i}; \boldsymbol\theta))\right]^\top\R^{-1}\left[\y(t_i) - H({\xa}(t_{i}; \boldsymbol\theta))\right],\nonumber\\
    {\xa}(t_{i}; \boldsymbol\theta) &= {\xa}^{\text{,ML}}(\xb(t_{i}), \y(t_{i}); \boldsymbol\theta).
\end{align}
The same idea can be applied to 4DVar.

\subsection{Assimilation in a latent space}

Recall that autoencoders (see Chapter 1) learn a low-dimensional latent space for a given dataset. An active area of research has been to perform DA in such a latent space \citep{melinc_3d-var_2024,pasmans_ensemble_2025}, for high-dimensional systems such as the atmosphere. One might want to assimilate in the latent space for two reasons:
\begin{enumerate}
    \item DA is costly in high dimensions. If the assimilation step can be performed in a lower-dimensional latent space, the computational cost can be significantly reduced.
    \item DA methods, as discussed in Section~\ref{sec:da}, typically make Gaussian assumptions. Variational autoencoders impose Gaussian structure on the latent space, so Gaussian DA methods can potentially be more accurately applied here than in a non-Gaussian full space.
\end{enumerate}

Suppose one has trained an autoencoder on time-series of the system of choice, resulting in an encoder $E$ and decoder $D$. Then, to perform assimilation in the latent space at time $t_i$, we compute the latent space representation of the forecast as
\begin{equation*}
    \mathbf{z}^\text{f}(t_i) = E(\xf(t_i)).
\end{equation*}
For an ensemble forecast, the encoder would be applied to each ensemble member, obtaining a latent space ensemble $\{\mathbf{z}^\text{f}_j(t_i)\}_{j=1}^m$. We then define the latent space observation operator to be
\begin{equation*}
    H' = H \circ D;
\end{equation*}
that is, to get from a point in the latent space to one in the observation space, one first applies the decoder to return to the full space, and then the observation operator.

Now, one can apply a DA algorithm to the latent space forecast $\mathbf{z}^\text{f}(t_i)$ and observation $\y(t_i)$, with observation operator $H'$, to obtain the latent space analysis $\mathbf{z}^\text{a}(t_i)$. To obtain a full space analysis one can apply the decoder,
\begin{equation*}
    \xa(t_i) = D(\mathbf{z}^\text{a}(t_i)).
\end{equation*}

We have not yet addressed how to obtain the latent space forecast covariance matrix. In the case of the EnKF, this can be directly estimated from the latent space ensemble. For 3DVar, the background covariance can be estimated from consecutive latent space states \citep{melinc_3d-var_2024}.

\subsection{Generative data assimilation}

We have thus far focused on the problem of state estimation: obtaining a good point estimate of the system state at a given time. Filtering, as discussed in Section~\ref{sec:da}, is the problem of instead estimating a probability distribution over possible states. Specifically, one wants to estimate the \emph{filtering distribution},
\begin{equation*}
    \mathbb{P}(\xt(t_i)|\{\y(t_j)_{j=0}^i\});
\end{equation*}
that is, the conditional probability of the state at $t_i$ given all the observations from time $t_0$ to time $t_i$.

An emerging area of research applying ML to DA is to approximately sample this distribution using generative ML. Specifically, several approaches have used diffusion models as priors for generative DA \citep{rozet_score-based_2023,bao_score-based_2024,yang_generative_2025}. Diffusion models are discussed in Chapter 6.

One approach is to first obtain a generative model that can approximately sample from the probability distribution $\mathbb{P}(\xt)$ by training on either simulated time-series or reanalysis. Once this unconditional model is trained, there are various approaches to sample from the approximate distribution conditioned on observations. While such methods seem promising, the issue of how to perform conditional sampling with diffusion models in a principled and tractable way is a difficult one \citep{zhao_conditional_2025}. Methods for conditioning that have been used for generative DA include diffusion posterior sampling \citep{rozet_score-based_2023} and CoPaint \citep{yang_generative_2025}.

Another approach is to directly train a conditional diffusion model, with the disadvantage that the trained model is specific to the observation network used in training. See \cite{hodyss_using_2026} for a review of these approaches.

\subsection{End-to-end approaches}\label{sec:end-to-end}

\emph{End-to-end} approaches attempt to predict future observations given past observations. These methods no longer distinguish between forecast and analysis steps, dispensing with a physics-based model as well as traditional DA.

The idea is to train a model $\y^{\text{ML}}$ that takes as input some number $M$ of past observations, $\{\y(t_j)\}_{j=i-M}^{i-1}$, and output a predicted observation at time $t_i$.

Such a model can be trained using a loss function such as
\begin{equation}
    J(\boldsymbol\theta) = \frac{1}{n - M+1}\sum_{i=M}^n\|\y(t_i) - \y^{\text{ML}}(\{\y(t_j)\}_{j=i-M}^{i-1}; \boldsymbol\theta)\|^2;
\end{equation}
we now no longer rely on a physics-based model nor reanalysis.

One of the advantages of this approach is that, by training on observations, one can avoid the model biases present in reanalysis. The cost of training on raw observations, however, is that one discards much of the important information contained within the reanalysis, which functions as a physics-based interpolator between sparse and noisy observations; see section \ref{ssec:reanalysis}. Furthermore, the predicted observations may themselves be sparse and indirect.

Recent successful examples of this approach include \cite{alexe_graphdop_2024,allen_end--end_2025}. It should be noted, however, that the latter was trained first using reanalysis and then ``fine-tuned'' using observations. Purely observation-based approaches, such as \citet{alexe_graphdop_2024}, have thus far not reached the skill of leading physics-based models nor ML models trained on reanalysis.



\bibliographystyle{abbrvnat}
\bibliography{references}

\end{document}